\def\gore{W. curdt et al.: The coronal convection}

\documentclass{ceab}  
\usepackage{epsfig}

\begin{document}

\title{the coronal convection}

\author{\normalsize W. CURDT$^1$, H. TIAN$^{1,2}$, E. MARSCH$^1$ \vspace{2mm} \\
        $^1$\it Max Planck Institute for Solar System Research \\
            \it Max-Planck-Str. 2, 37191 Katlenburg-Lindau, Germany\\
        $^2$\it School of Earth and Space Sciences, Peking University\\
            \it Beijing 100871, China}
\maketitle

\begin{abstract}
We study the hydrogen Lyman emission in various solar features -- now including
Ly-$\alpha$ observations free from geocoronal absorption -- and investigate
statistically the imprint of flows and of the magnetic field on the line profile and radiance
distribution. As a new result, we found that in Ly-$\alpha$ rasters locations with
higher opacity cluster in the cell interior, while the network has a trend
to flatter profiles. Even deeper self reversals and larger peak distances were
found in coronal hole spectra. We also compare simultaneous Ly-$\alpha$ and Ly-$\beta$
profiles. There is an obvious correspondence between asymmetry and redshift
for both lines, but, most surprisingly, the asymmetries of Ly-$\alpha$ and Ly-$\beta$
are opposite. We conclude that in both cases downflows determine the line profile,
in case of Ly-$\alpha$ by absorption and in the case of Ly-$\beta$ by emission.
Our results show that the magnetically
structured atmosphere plays a dominating role in the line formation and indicate
the presence of a persisting downflow at both footpoints of closed loops.
We claim that this is the manifestation of a fundamental mass transportation
process, which Foukal back in 1978 introduced as the 'coronal convection'.

\end{abstract}

\keywords{mass transport - chromosphere - line profile - Lyman alpha} 

\section{Introduction}
\large

The basic idea of this communication is to show that the well-known net redshift
of transition region (TR) emission and the new observations of the Ly-$\alpha$ and
Ly-$\beta$ profiles obtained by SOHO-SUMER are different manifestations of the
same fundamental massflow process.

It is known since long that in TR emission all lines appear with
a net redshift, which peaks at a temperature around log\,$T$/K=5 (e.g., Brekke 1997).
Dammasch et al. (2008) have demonstrated that this redshift can be explained by
the fact that closed loops higher up in the TR have both footpoints
redshifted. Such coronal downflows have also been reported by Hinode (Tripathi 2009,
del Zanna 2008). Already 30 years ago Foukal (1978) mentioned such flows and introduced
the term 'Coronal Convection'.
The common understanding of these observations is that in TR
loops that are closed, material is draining down on both sides.
Near their footpoints, the loops are not only redshifted, but they are
also brighter than elsewhere, and this brightness-to-redshift relationship
easily explains the net redshift of TR emission and the redshift in the network
contrast (Curdt et al. 2008a). We show that clear signatures of dynamical
processes are also seen in the optically thick lines of the hydrogen Lyman series.

\section{Recent Ly-$\alpha$ observations}

The entire Lyman series and the long-wavelength portion of the
hydrogen recombination continuum fall into the spectral range of the SUMER
spectrometer. Ly-$\alpha$ -- by far the most important line and carrying
75\,\% of the spectral radiance in this range -- plays an important role for
the radiative energy transport in the solar atmosphere. Despite of this
importance, however, such observations were not completed,
because the bright Ly-$\alpha$ line would saturate the detector.
Recently the SUMER team started unconventional observations, where 80\,\% of
the aperture is vignetted by the partially closed door.
Several of these rasters were completed at various $\mu$ angles in summer
2008, when the Sun was very quiet. In April 2009 also coronal hole (CH) data was
obtained. Later on, prominence, filament and sunspot targets were selected.
In some of the rasters, the wavelength setting was within 23.3\,s switched back
and forth between Ly-$\alpha$ and Ly-$\beta$, thus allowing quasi-simultaneous
exposures for those lines (for observational details see Curdt et al. 2008b,
2010a, 2010b, Tian et al. 2009a, 2009b).
Some of the results that are relevant for our topic are described below.

\subsection{How Flows Affect the Profiles of Ly-$\alpha$ and Ly-$\beta$}

In all exposures we also recorded a TR line as line-of-sight (LOS) velocity indicator.
In the case of Ly-$\alpha$,
the Doppler flow in each pixel has been determined by the shift of the
$\lambda$\,1206\,Si\,{\sc{iii}} line centroid.
\begin{figure}[h] 
\begin{center}
  \epsfig {file=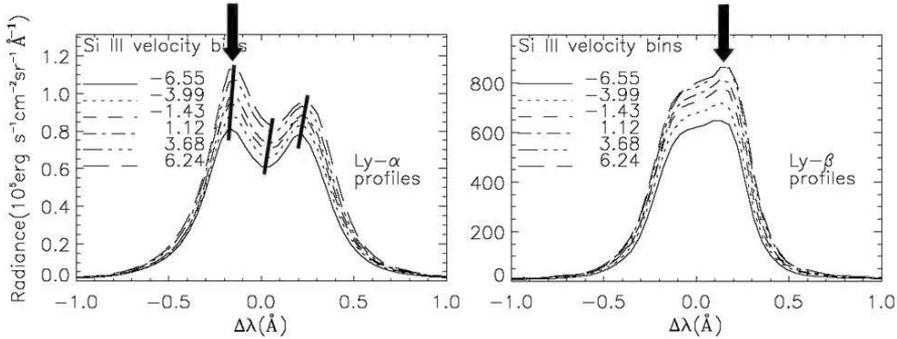,width=12cm}
  \caption {Left panel: Line profiles of Ly-$\alpha$
  observed near disk center. We show the profiles for six equally spaced velocity
  bins, which were defined by a pixel-by-pixel comparison of the
  $\lambda$ 1206 Si\,{\sc{iii}} line centroids with the rest wavelength.
  Negative values correspond to upflows, positive values to downflows.
  There is a clear correspondence between asymmetry and downflows.
  Right panel: Similarly, we analysed the line profiles of Ly-$\beta$.
  It is most obvious that the asymmetries in the Ly-$\alpha$ (left) and the Ly-$\beta$
  lines (right) are reversed, and there is a clear correspondence
  between asymmetry and downflows for both lines.}
\end{center}
\end{figure}
We sort the Ly-$\alpha$ profiles by the Doppler flow in each pixel
and define six equally spaced velocity bins. The different profiles of each bin are
displayed in Fig.\,1a. There is an obvious and remarkable result:
The blue horn is always stronger than the red horn, and the asymmetry
increases significantly with the redshift.
Both peaks and the central reversal are offset towards the red with
increasing downflow, as indicated by tracing lines.
Similarly, we display in Fig.\,1b the profiles of Ly-$\beta$.
Also for Ly-$\beta$ there is an obvious correspondence between asymmetry and
redshift and between asymmetry and radiance. But most surprisingly the asymmetries
of Ly-$\alpha$ and Ly-$\beta$ are reversed.
It is well known that higher order Lyman lines become more optically thin.
Our results clearly show that the trend towards self-reversed, red-horn dominated profiles
is discontinued for Ly-$\alpha$.
Interestingly, this has already been noticed three decades ago in OSO\,8 data (Gouttebroze et al. 1978).
They clearly show in their Fig.\,1 the red-horn dominance in Ly-$\beta$ and the blue-horn
dominance of Ly-$\alpha$. But they don't give any explanation other than:
'The asymmetries are variable with time and location and are probably related
to large-scale motions of the atmosphere.' We now arrived at a heuristic model
to explain the opposite asymmetries, where downflows and
opacity are dominating the profile, but play a different role.
\begin{figure} 
\begin{center}
  \epsfig {file=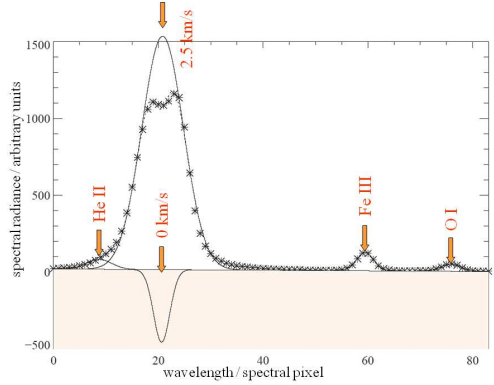,width=7cm}
  \caption {5-component multigaussfit, assuming a negative Gaussian to account for the absorption.
  The best fit was achieved for a Ly-$\beta$ line that is redshifted by 2-3 km/s
  relative to the absorption at zero velocity.}
\end{center}
\end{figure}
Both lines are optically thick, but the opacity in Ly$-\beta$ is much less than that in Ly$-\alpha$.
In Fig.\,2 we show the average line profile
of several chromospheric lines around Ly-$\beta$ obtained in a quiet Sun (QS) location.
We adopted the wavelength values of the SUMER spectral atlas (Curdt et al. 2001) and
completed a precision 5-component multigaussfit, assuming a negative
Gaussian to account for the absorption. The fit with the least errors was achieved,
if we assume the absorption to occur at zero velocity and the full curve
at a redshift of 2 to 3 km/s relative to this. This suggests that
Ly-$\beta$ (and the higher-order Lyman lines) still behave like
typical TR lines, even in Ly-$\beta$ the redshifted footpoints are shining through.
For Ly-$\alpha$ things change, the opacity of Ly$-\alpha$ is so high,
that any directional information is lost at $\tau=1$ because of the partial redistribution
process. In contrast to Ly$-\beta$ and all the higher Lyman lines, which are still
dominated by redshifted {\it emission} (which narrows the blue peak),
Ly$-\alpha$ is dominated by redshifted {\it absorption} (which suppresses the red peak).
Below we will give additional proof for this interpretation.
\begin{figure}[t] 
\begin{center}
  \epsfig{file=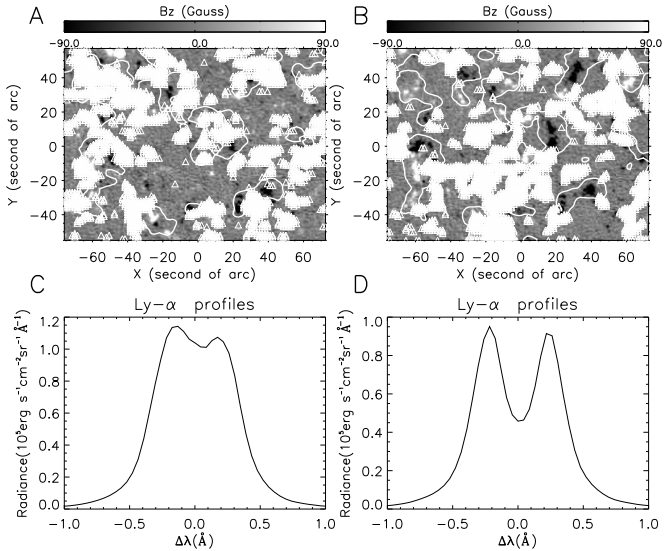, width=9cm}
  \caption {Magnetograms in grey, overlaid with Ly-$\alpha$ brightness contours (white).
  Right: Pixels with the deepest self-reversal are marked. Left: Pixels with the
  flattest profiles are marked 
  (cf. for details, Tian 2009a).}
\end{center}
\end{figure}
\subsection{How the Magnetic Field Affects the Profiles of Ly-$\alpha$}

In Fig.\,3 we show a MDI magnetogram of a QS area near disk center
The overlaid contours indicate enhanced Ly-$\alpha$ radiance and show the top-25\,\% level.
Here we sort the profiles by the relative depth of their self-reversal. In panel A we marked pixels
out of the bottom-25\,\% regime, areas with flatter profiles (reduced opacity)
obviously cluster along the network lane.
In contrast, pixels from the top-25\,\% regime - shown in panel B - cluster in
the cell interior. This suggests a lower opacity above the magnetic network as
compared to the low-lying loops in the cell interior.
We assume that in the network magnetic loops open into over-arching structures,
while the cell interiors are filled with low-lying loops.
Less self-reversed profiles cluster along the network boundary,
here the expanding magnetic field forms funnel-like structures
with reduced opacity. In other words, chromosphere and transition
region are extended above the network.

\subsection {Lyman Emission in Coronal Holes}

The arguments raised so far imply that in CHs the
effects should be even more expressed. Indeed, in the CH we found the
largest peak separation in the Ly-$\alpha$ profile
and the deepest self-reversals in the Ly-$\beta$ profile, in both cases
indication of extremely high opacity (cf., Tian et al. 2009b).

In Fig.\,4 we show the profiles of Ly-$\alpha$ (left panel) and of Ly-$\beta$
as observed in a polar CH. It is most interesting that the Ly-$\beta$
profiles in the polar CH have a blue peak dominance similar to Ly-$\alpha$.
Xia (2003) mentioned indications of such a behaviour in equatorial CHs.
 In our work we find very clearly that
most Ly-$\beta$ profiles are stronger in the blue peak in the polar CH. Ly-$\beta$ now starts
to behave like Ly-$\alpha$, and this in line with the argument used before and
proves that our interpretation is consistent.
This difference in behaviour suggests that the CH is void of overarching loops.
Here, the chromosphere is filled everywhere with low-lying loops,
that are observed at an orthogonal view.
\begin{figure}[h] 
\begin{center}
  \epsfig {file=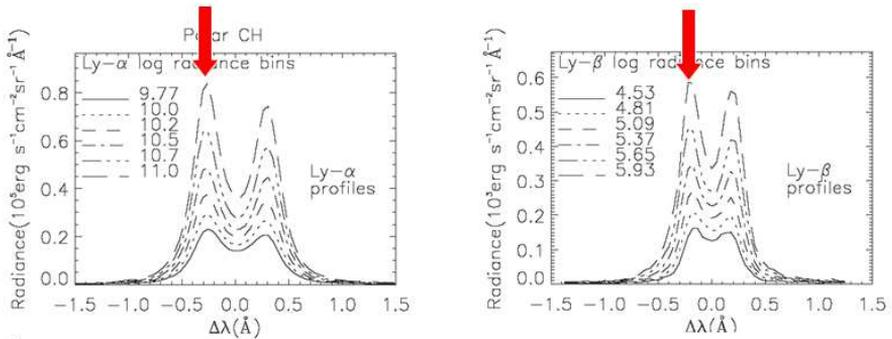,width=12cm}
  \caption {Ly-$\alpha$ and Ly-$\beta$ averaged profiles in a polar CH, the profiles
are sorted into six radiance bin. For both lines
the peak separation increases significantly, and in Ly-$\beta$ the
red-horn dominance - common in the QS - changes to blue-horn dominance.}
\end{center}
\end{figure}
\section {Red and Blue Branch of the Coronal Convection}

In the previous sections we have shown the signature of downflows in
optically thin and in optically thick lines and that there is evidence for
persistent downflows in closed loops. In Foukal's notion this would be the
'red branch' of the coronal convection. The blue branch is less clear.
There is evidence for continuous upflow in magnetically open structures
from the spectrometers on SOHO and Hinode. Here, we refer to the work
of Wiegelmann et al. (2005). In their Fig.\,2 (not reproduced here)
open field lines are shown in brown, closed ones in yellow.
Indeed a correlation exists between plasma outflow (shown in blue) and
places with open field lines., indicative of a quasi-continuous process that
is guided by magnetic field lines (Marsch et al. 2008).
Also transient scenarios like jets or spicular activity have been suggested
by the community to be responsible for the material upflow.
This is not necessary in contradiction with the open-fieldline approach.
But at this time a clear answer can not be given.
\begin{figure} 
\begin{center}
  \epsfig {file=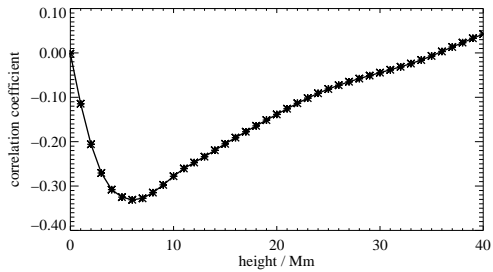,width=7cm}
  \caption {Correlation of the LOS-component of the magnetic field
(as extrapolated from MDI data) and the reversal depth of Ly-$\alpha$.}
\end{center}
\end{figure}

\section{Ly-$\alpha$ profiles and magnetic field orientation}

Gun\'ar et al. (2008) claim that the orientation of a prominence axis relative to the
LOS has an imprint on the self-reversal depth of Ly-$\alpha$, such that
the reversal is deeper, when the magnetic field is oriented vertical to the LOS.
The theoretical work is based on Heinzel et al. (2001, 2005).
And observational evidence has recently been reported by Schmieder et al. (2007) and
Curdt et al. (2010a).

We correlated the LOS-component of the
extrapolated magnetic field in a QS area at different coronal heights versus
the reversal depth the Ly-$\alpha$ profile.
Fig.\,5  shows a small, but clear anti-correlation, which has its minimum around
6 \,Mm. This suggests that the orientation of the magnetic field relative to the LOS
has an imprint on the source function and plays an important role not only
in a prominence, but does it everywhere.

\section{Conclusion}

Despite the importance of Ly-$\alpha$ for radiative transport and many
other processes in the solar atmosphere, there is a clear disproportion
if one compares Ly-$\alpha$ focussed articles to publications on optically thin
lines. We presented a set of empirical results, which we managed to explain
in a consistent manner and which resulted into a heuristic model, where
opacity effects and large-scale flows are the main elements.
We now started quantitative work along our heuristic model.

\section*{References}
\begin{itemize}
\small
\itemsep -3pt
\itemindent -20pt
\item[] Brekke, P., Hassler, D. M., Wilhelm, K.: 1997, $Sol.Phys.$ {\bf 175}, 349.
\item[] Curdt, W. {\it et al.}: 2001, $A\&A$ {\bf 375}, 591.
\item[] Curdt, W., Tian, H., Teriaca, L., Sch\"uhle, U.: 2010a, $A\&A$ {\bf 511}, L4. 
\item[] Curdt, W., Tian, H., B.N. Dwivedi, Marsch, E.: 2008a, $A\&A$ {\bf 491}, L13.
\item[] Curdt, W., Tian, H., Teriaca, L., Sch\"uhle, U., Lemaire, P.: 2008b, $A\&A$ {\bf 492}, L9.
\item[] Curdt, W., Tian, H.: 2010b, $ASPC$ {\bf 428}, 81.
\item[] Dammasch, I.E., Curdt, W., Dwivedi, B.N., Parenti, S.: 2008, $Ann. Geophys.$ {\bf 26}, 2955.
\item[] del Zanna, G.: 2008, $A\&A$ {\bf 481} L49.
\item[] Foukal, P.: 1978, $ApJ$ {\bf 223}, 1046.
\item[] Gouttebroze, P., Lemaire, P., Vial, J.-C., Artzner, G.: 1978, $ApJ$ {\bf 225}, 655.
\item[] Gun\'ar, S., Heinzel, P., Anzer, U., Schmieder, B.: 2008, $A\&A$ {\bf 490}, 307.
\item[] Heinzel, P., Anzer, U.: 2001, $A\&A$ {\bf 375}, 1082.
\item[] Heinzel, P., Anzer, U., Gun\'ar, S.: 2005, $A\&A$ {\bf 442}, 331.
\item[] Marsch, E., Tian, H., Sun, J., Curdt, W., Wiegelmann, T.: 2008, $ApJ$ {\bf 685}, 1262.
\item[] Schmieder. B., Gun\'ar, S., Heinzel, P., Anzer, U.: 2007, $Sol.Phys.$ {\bf 241}, 53.
\item[] Tian, H., Curdt, W., Marsch, E., Sch\"uhle, U.: 2009a, $A\&A$ {\bf 504}, 239.
\item[] Tian, H., Teriaca, L., Curdt, W., Vial, J.-C.: 2009b, $ApJ$ {\bf 703}, L152. 
\item[] Tripathi, D., Mason, H.E., Dwivedi, B.N., del Zanna, G.: 2009, $ApJ$ {\bf 694}, 1256.
\item[] Wiegelmann, T., Xia, L.-D., Marsch, E.: 2005, $A\&A$ {\bf 432}, L1.
\item[] Xia, L.-D.: 2003, $PhD-thesis$, Georg-August-Universit\"at G\"ottingen.

\end{itemize}

\end{document}